\documentclass[a4paper,10pt,twoside]{cpc-hepnp}

\usepackage{multicol}
\usepackage{graphicx}
\usepackage{booktabs}
\usepackage{amssymb,bm,mathrsfs,bbm,amscd}
\usepackage[tbtags]{amsmath}
\usepackage{lastpage}
\usepackage{amsfonts}
\usepackage{amssymb,epsfig}
\usepackage{amsmath}
\usepackage{array}
\usepackage{subfig}
\usepackage[T1]{fontenc}
\usepackage[all]{xy}
\usepackage{epsfig,psfrag}
\usepackage{rotating}
\usepackage{ragged2e}
\usepackage{float}
\usepackage{placeins}
\usepackage{appendix}
\usepackage{makecell}
\allowdisplaybreaks
\makeatletter
\AtBeginDocument{%
  \expandafter\renewcommand\expandafter\subsection\expandafter
    {\expandafter\@fb@secFB\subsection}%
  \newcommand\@fb@secFB{\FloatBarrier
    \gdef\@fb@afterHHook{\@fb@topbarrier \gdef\@fb@afterHHook{}}}%
  \g@addto@macro\@afterheading{\@fb@afterHHook}%
  \gdef\@fb@afterHHook{}%
}
\makeatother

\def\be{\begin{equation}}
\def\ee{\end{equation}}
\def\zz{z,\hat{z}}

\begin{document}



\title{Superpotentials and Geometric Invariants of Parallel/Complete Coincident/Part Coincident D-brane System on Compact Calabi-Yau Manifold}

\author{%
\quad Fei Li$^{1}$
\quad Fu-Zhong Yang$^{1}$\email{fzyang@ucas.ac.cn}%
}
\maketitle

\address{%
$^1$ University of Chinese Academy of Sciences, \\No.19(A) Yuquan Road, Shijingshan District, Beijing, P.R.China\\
}

\begin{abstract}
For D-brane system with three D-branes on compact Calabi-Yau threefolds, the dual F-theory fourfolds for parallel/complete coincident/part coincident D-brane system is constructed by the type II/F-theory duality. Complete coincident means that the three D-branes coincide and part coincident represents the coincident of two of the three D branes. The low energy effective superpotentials are calculated by mirror symmetry, GKZ-system method and the type II/F-theory duality on the B-model side, respectively. Using the mirror symmetry, A-model superpotentials and the Ooguri-Vafa invariants are obtained from the B-model side. These results indicate that the superpotential contributed by one of three parallel D-branes is identical with the D-brane system with only one D-brane, which is a signal of decoupling of the parallel topological D-branes. However, the superpotential and Oogrui-Vafa invariants are different among parallel, complete coincident and part coincident D-brane system, which show the evidence of the phase transition due to the enhanced gauge symmetry in the low energy theories and the geometrical singularity.
\end{abstract}




\begin{multicols}{2}

\section{Introduction}

\label{sec:intro}
 In the $N=2$ supersymmetric theories, the closed-string mirror symmetry gives an equivalence between A-model parameterized by K\"{a}hler moduli in the terms of the quantum geometry and B-model parameterized by complex structure moduli in terms of the classical geometry. Mirror symmetry provides many techniques for the variation of physical structures over their moduli spaces\cite{CQ.14,Ba.18}, and in the closed-string sector\cite{LVW.89,BCOV.94} has a relatively perfect solution in the early works. With the appearance of D-branes, we take more attention to the open-string sector in recent years\cite{SU.15}. The supersymmetry breaks into $N=1$ when D-brane is included and that leads the application of open-closed mirror symmetry\cite{MK.94,MP.01,JW.06}, e.g., the quantum corrected domain wall tensions on the Calabi-Yau threefolds can be calculated with open-closed mirror for compact Calabi-Yau manifolds.\\
 The D-brane superpotential is a section of a special holomorphic line bundles of the moduli space from the mathematical perspective. It is defined as the F-term of low-energy effective theory and determines the string vacuum structure from the physical perspective. The superpotential is the generating function of all disk instantons from the worldsheet point of view, and it can be calculated by relative period. The D-branes on A-model, which are called A-branes, wrap on the special Lgarangian submanifolds of Calabi-Yau manifold $W_3$. Correspondingly, B-branes wrap on holomorphic submanifolds of Calabi-Yau manifolds $M_3$. The expression of instanton expansion of superpotential on the A-model side encodes the number of BPS states which corresponds to the Ooguri-Vafa invariants mathematically\cite{AV.00}. These invariants give an important mathematical language which has not been studied systematically in theory so far. It is hard to calculate in the A-model because of the non-perturbed quantum corrections. Nevertheless, we can figure it out in B-model, and then mirror it to A-model. For non-compact Calabi-Yau manifolds, several methods that can be used to calculate the D-brane effective superpotential are localization\cite{AV.00,AKV.02,A.05}, topological vertex and direct integration related to $N=1$ special geometry\cite{LMW.02.1,LMW.02.2}. Further, for compact Calabi-Yau manifolds, some techniques have been evolved, e.g., mixed Hodge structure variation, Gauss-Manin connection\cite{JS.08,JS.09}, the blow-up method\cite{GHKK.09} and the GKZ-generalized hypergeometric system for open-closed sectors. This paper we focus on is to calculate D-brane superpotentials for compact Calabi-Yau manifolds with open-closed mirror symmetry and generalized GZK system\cite{JXT.17,XFJ.14.01,XFJ.14.02,CS.14,ZSS.15,ZH.16,AHMM.09,LS.09}.\\
 The dual description of the superpotential in Type-II string theory can be found in F-theory. To be precise, the D-brane superpotentials in the string theory of Type-II is dual to the background flow in F-theory\cite{AHJMMS.09,JMW.09}. This duality provides us with a method to calculate the D-brane superpotentials in Type-II string theory. Some works which are related to the calculation of the superpotentials near the limit point and the invariants for the system with two D-branes have been done\cite{JXT.17}. And it is mentioned that the parallel and the coincident D-branes phases correspond to the Coulomb branch and the Higgs branch of the non-Abelian gauge theory on the worldvolume of D-brane system respectively. Motivated and guided by the work, we calculate and compare the system with three D-branes. The system with three branes are more generalized than two branes, and it complicates the research. The coincident D-branes phases are divided into complete coincident phase and part coincident phase. Complete coincident means that the three D-branes coincide and part coincident
represents the coincident of two of the three D branes. \\
The organization is as follow. In section $2$, we review the background related to physics and mathematics. This section gives an overview of toric brane geometry and GKZ system. In section $3$, we concentrate on two models, D-brane system on the mirror quintic and on hypersurface $X_8(1,1,2,2,2)$. The superpotentials of parallel, complete coincident and part coincident D-brane phase are calculated and discussed from each model. The Ooguri-Vafa invariants are extracted in different D-branes models as well. First, we calculate superpotentials and invariants of the parallel D-brane phase (Coulomb phase) with three parallel branes. Second, we calculate the superpotentials and invariants of complete coincident D-brane phase (Higgs phase) in which three parallel D-branes coincide. Third, we calculate the superpotential of part coincident D-brane phase (Coulomb phase-Higgs phase) in which two of the three parallel D-branes coincide. The last section is a brief summary.

\section{Toric brane geometry and GKZ system}

\subsection{Mirror symmetry}
Mirror symmetry is the duality between Type II A  and Type II B topological string theory. The corresponding D-branes are A-brane and B-brane, respectively. The supersymmetry breaks from $N=2$ to $N=1$ when D-branes are included. Sometimes we also call $N=1$ mirror symmetry as open-closed mirror symmetry since this involves the open and closed part.\\
Closed mirror symmetry is more fundamental than open-closed mirror symmetry. Taking the brane to infinity and we just suppose there is an infinity, then the open-closed mirror symmetry degenerates to the closed mirror symmetry. It is necessary to note is that there are two ways to explain the open-closed. One way to understand is that an open string falls on the brane and another one on the anti-brane (the brane and anti-brane can be regards as two branes that carry opposite charges), then two discs was glued to be a sphere, which is closed string. Another way to understand is that closed string does not needs brane since brane in string theory is just a boundary condition. Closed string is a Rieman surface without boundary, which can be projected into Calabi-Yau manifold, while open string is Rieman surface with boundary.\\
The A-brane is wrapped on the special Lagrangian submanifold $L$ of the Calabi-Yau manifold $W$, while the B-brane is wrapped on the holomorphic submanifold $\mathcal{D}$ of the Calabi-Yau manifold $M$. Mirrored symmetry with D-brane has a more precise description in mathematics: The equivalence between the derived category of coherent sheaves on a Calabi-Yau manifold and the Fukaya category of its mirror. $N=1$ Mirror symmetry contacted two completely different D-brane geometries $(W, L)$ and $(M, D)$.

\subsection{The effective superpotentials in Type-II string theory and F-theory}
On the B-model side, the space-filling D5-branes wrap on an reducible curve $C=\sum_{i}\mathcal{C}_i$ and C embedded in a divisor $\mathcal{D}$ of Calabi-Yau 3-fold $M_3$. The effective superpotential is:
\be
\mathcal{W}_{\mathcal{N}=1}(z,\hat{z})=\Pi_\gamma(z,\hat{z})=\int_\gamma\Omega^{(3,0)}(z,\hat{z}),~~~\gamma\in H_3(M_3,\mathcal{D})
\ee
The superpotential can be written as a linear combination of relative period\cite{LMW.02.1,LMW.02.2}:
\be
\begin{split}
\mathcal{W}_{\mathcal{N}=1}(z,\hat{z})&=\sum N_{\alpha}\Pi_{\alpha}(z,\hat{z})\\
&=\mathcal{W}_{open}(z,\hat{z})+\mathcal{W}_{closed}(z)\\
&=\mathcal{W}_{brane}(z,\hat{z})+ \mathcal{W}_{flux}(z),\\
\Pi_\alpha(z,\hat{z})&=\int_{\alpha}\Omega^{(3,0)}(z,\hat{z}).
\end{split}
\ee
On the A-model side, a general form of D-brane superpotential is:
\be
\begin{split}
\mathcal{W}_{\mathcal{N}=1}=&\mathcal{W}_{classical}(t,\hat{t})+\mathcal{W}_{instanton}(q,\hat{q})\\
=&(\frac{1}{2}\mathcal{K}_{jk}t^{j}\hat{t}^{k}+b)+\sum_{k,m}G_{k,m}q^{k}\hat{q}^{m}\\
=&(\frac{1}{2}\mathcal{K}_{jk}t^{j}\hat{t}^{k}+b)+\sum_{n}\sum_{k,m}\frac{N_{k,m}}{{n^{2}}q^{nk}\hat{q}^{nm}},
\end{split}
\ee
where $t^{j}$ and $\hat{t}^{k}$ are the closed and open K\"{a}hler moduli respectively, $q=exp(2\pi it)$, $\hat{q}=exp(2\pi i\hat{t})$. $\{G_{k,m}\}$ are the open Gromov-Witten invariants labeled by relative homology class. $\vec{m}$ represent the elements of $H_{1}(L)$, $\vec{k}$ represent the elements of $H_{2}(W_{3})$, and $\mathcal{K}_{jk}$ is the combination coefficient. $\{N_{\vec{k},\vec{m}}\}$ are the Ooguri-Vafa invariants.\\
There is a duality between the type II string theory with D-brane systems on complex Calabi-Yau threefold $M_{3}$ and the F-theory compactified on the Calabi-Yau fourfold $M_{4}$ without any branes but with fluxes.\\
The superpotential of 4-form flux $G_{4}$ in F-theory compactified on the Calabi-Yau 4-fold $M_{4}$ is a section of the Hodge line bundle in the complex structure moduli space $\mathcal{M}_{CS}(M_4)$. This superpotnetial is called Gukov-Vafa-Witten superpotential\cite{GVW.99,AHJMMS.09}:
\be
\begin{split}
\mathcal{W}_{GVW}(M_4)=&\int_{M_4}G_4\wedge\Omega^{(4,0)}\\
=&\sum_\Sigma N_\Sigma(G_4)\Pi_\Sigma(\zz)+\mathcal{O}(g_s)+\mathcal{O}(e^{-1/g_s}),
\end{split}
\ee
The leading term of above equation on the right-hand side is the D-brane superpotential $\mathcal{W}_{N=1}$ and $g_{s}$ is the string coupling strength.\\
There is a duality between the periods of holomorphic $(4,0)$ form on the non-compact 4-fold $\tilde{M}_{4}$ and the relative periods of the brane geometry ($M_{3}$, $\mathcal{D}$). From mirror symmetry, one obtains the relation between the different compactifications:
\be
\xymatrix{
\multirowcell{8}{(W_{3},L)\\(A-branes)}\ar[d]_{4f}^{dual} \ar@{<->}[rr]_{symmetry}^{mirror}
                         & &\multirowcell{8}{(M_{3},\mathcal{D})\\(B-branes)} \ar[d]_{4f}^{dual}\\
\tilde{W}_{4} \ar@{<->}[rr]^{mirror}_{symmetry}  & &\tilde{M}_{4} }
\ee
It is important that the moduli space $\mathcal{M}(M_{3},\mathcal{D})$ of open-closed system and the moduli space $\mathcal{M}(\tilde{M}_{4})$ of non-compact 4-fold $\tilde{M}_{4}$ are isomorphic. Thus the Picard-Fuchs equation can be used to calculate the superpotential by open-closed duality.
Further, the structure of the mirror pair $(\tilde{W}_{4},\tilde{M}_{4})$ is as follows:
\be
\xymatrix@C=1.5cm{
W_{3} \ar[r]   &\tilde{W}_{4} \ar@{<->}[r]_{symmetry}^{mirror} \ar[d]_{\pi}      &\tilde{M}_{4} \ar[d]_{\pi}  &\tilde{\varepsilon} \ar[l] \\
&T        &M_{3}  }
\ee
Where $\tilde{M}_{4}$ is an elliptic fibration on a 3-fold $M_{3}$, while its mirror partner $\tilde{W}_{4}$ is a fibration with a Calabi-Yau 3-fold $W_{3}$ as fiber and a disk T as the base.\\
The mirror pair of four-fold $(\tilde{W}_{4},\tilde{M}_{4})$ are non-compact. To get the honest 4-dimensional F-theory compactification, a $P^{1}$ compactification of the non-compact base T of $\tilde{W}_{4}$ as follows:
\be
\xymatrix@C=1.5cm{
W_{3} \ar[r]    &W_{4} \ar@{<->}[r]_{symmetry}^{mirror} \ar[d]_{\pi}        &M_{4}  \\
&P^{1}     &(F-theory)}
\ee
In this way one obtains a mirror pair of compact Calabi-Yau 4-fold $(W_{4}, M_{4})$, where $M_{4}$ is the 4-fold for F-theory compactification and duel to the B-brane geometry $(M_{3}, \mathcal{D})$.
The large volume limit $Im S=Vol(P^{1})\to \infty$ maps under mirror symmetry to a weak coupling limit $g_{s} \to 0$. It is following in the figure:
\be
\xymatrix@C=2cm{
W_{4} \ar@{<->}[r]_{symmetry}^{mirror}  \ar[d]_{Vol(P^{1})\to \infty}       &M_{4}  \ar[d]^{g_{s} \to 0}   \\
\mathcal{W}_{4} \ar@{<->}[r]_{symmetry}^{mirror}                     & \mathcal{M}_{4}   }
\ee
In this limit, the D-brane superpotential $W_{N=1}$ is obtained from the GVW superpotential $W_{GVW}$ of F-theory as follows:
\be
\begin{split}
\mathcal{W}_{\mathcal{N}=1}(M_3,\mathcal{D})=&\sum N_\Sigma(G_4)\Pi_\Sigma(\zz)\\
=&\lim_{g_s\rightarrow 0}\mathcal{W}_{GVW}(M_4) ,
\end{split}
\ee
The zero-order term of the GVW superpotential and the D-brane superpotential $W_{N=1}$ is equal at the weak coupling limit. The moduli space $\mathcal{M}(\tilde{M}_{4})$ which is isomorphic to $\mathcal{M}(M_{3}, \mathcal{D})$ is a subspace of $\mathcal{M}(M_{4})$. The moduli space $\mathcal{M}(M_{4})$ is restricted to its subspace $\mathcal{M}(\tilde{M_{4}})$ by the weak coupling limit. Similarly, on the side of A model, the corresponding points in the moduli space $\mathcal{M}(W_{4})$ is the large base limit $Vol(P^{1})$\cite{BM.98,BM.05}.

\subsection{Toric geometry of D-branes system}
The Calabi-Yau manifolds in this paper is the hypersurfaces in ambient toric variety. We will use the method of toric polyhedral to define the GLSM for the mirror pairs of toric brane geometries\cite{HLY.96,HKTY.95.1,HKTY.95.2}.\\

\begin{center}
\begin{tabular}{ccc}
  Model:\quad &A-model \quad&B-model \\~\\
  Polyhedra:&$\nabla_{4}$ & $\Delta_{4}$ \\~\\
  Toric Variety:&$P_{\Sigma(\nabla_{4})}$ &$P_{\Sigma(\Delta_{4})}$ \\~\\
  Hypersurfaces(CY 3-fold):&$W_{3}$ &$M_{3}$
\end{tabular}
\end{center}

 For each pair of reflexive polyhedral $(\nabla_{4}, \Delta_{4})$ with a pair of complete fans $(\Sigma(\nabla_{4}),\Sigma(\Delta_{4}))$, there is a pair of toric varieties $(P_{\Sigma(\nabla_{4})},P_{\Sigma(\Delta_{4})})$. The defining polynomial for the hypersurface $M_{3}$ is:
\be
P=\sum^{p-1}_{i=0}a_{i}\prod_{k=1}^{4}X_{k}^{v_{i,k}^*},
\ee
Here $X_{k}$ are local coordinates on an open torus $(C^*)^{4}\subset P_{\Sigma(\Delta_{4})}$, and $v_{i,k}^{*}$ is the k-th coordinate of the integral point $v_{i}$ in $\nabla_{4}$. The coefficients $a_{i}$ are complex parameters related to the complex structure of $M_{3}$.\\
For the homogeneous coordinates $x_{j}$ on toric ambient space, the polynomial $P$ can be rewritten as
\be
P=\sum_{i=0}^{p-1}a_i\prod_{v\in\Delta_4}x_j^{\langle v,v_i^*\rangle+1}.
\ee
The $n$ parallel D-branes are defined by reducible divisor:
\be
\begin{split}
Q(\mathcal{D})=&\prod_{m=0}^n(\phi_ma_0+a_i\prod_{k=1}^4X_k^{v^*_{i,k}})\\
=&\sum_{k=0}^nb_k\prod_{v\in\Delta_4}x_j^{k\langle v,v^*_i\rangle+n},
\end{split}
\ee
The parallel D-brane geometry corresponds to the Coulomb phase of the gauge theory and the corresponding group is $U(1)\times U(1)\ldots U(1)$. The $U(1)$ group describes the electromagnetism that contains the Coulomb field.
The "enhanced polyhedron" expands the dimension of $v_{i}$ from four to five, and the relevant polyhedron consists of the extended vertices:
\be
\tilde{v}_j^*=(v_j^*,0)
\ee
The vertices of the parallel D-branes phase shaping the $\tilde{\nabla}_{5}$ are \cite{AHJMMS.12}
\be
\tilde{v}_j^*=\begin{cases}(v_j^*,0)& j=0,...,p-1,\\
(mv_i^*,1)& j=p+m, 0\leq m\leq n.\end{cases}
\ee
They define the non-compact 4-fold $\tilde{W}_4$. The graph is as follows
\begin{figure}[H]
\includegraphics[scale=0.4]{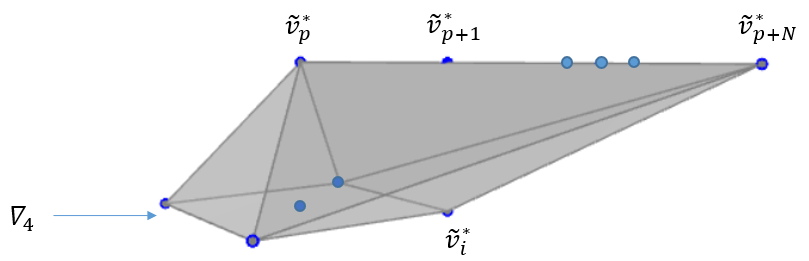}
\centering
\caption{Diagram: projected polyhedra of n parallel D-brane}
\end{figure}
\noindent The gauge group $U(1)\times U(1)\ldots U(1)$ is promoted to the $U(N)$ group when parallel D-branes coincide, while the phase is translated to the Higgs branch.
\begin{figure}[H]
\centering
\includegraphics[scale=0.4]{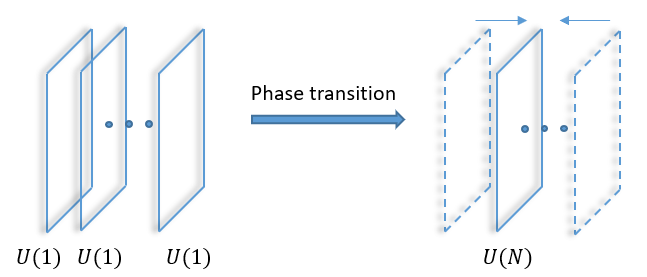}
\caption{phase transition}
\end{figure}
\noindent In toric geometry, the singular curve corresponds to the inner integral lattice point on one-dimensional boundary of the dual polyhedron. Each newly added point corresponds to the exceptional divisor in the blow-up of the Calabi-Yau manifold. Conversely, the removal of the inner point corresponds to the blow-down of these exceptional divisor\cite{BVS.95}. The $A_{n-1}$ singularity of the complex four-dimensional Calabi-Yau manifold is obtained by removing these vertices, which corresponds to the enhancement of the gauge symmetry group.
\begin{figure}[H]
\centering
\includegraphics[scale=0.4]{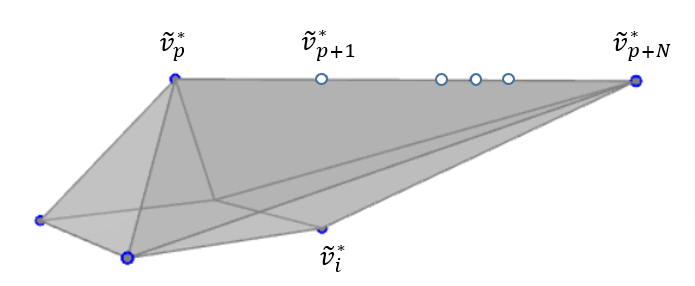}
\caption{Diagram: projected polyhedra of coincident D-brane}
\end{figure}
\noindent We have got the enhanced polyhedron which defines the non-compact four-fold $\tilde{X}_4^*$. Then the $P^{1}$ compactification can be obtained by adding the extra vertex. Combined with the compactifying point $\tilde{v}_{c}^{*}$, all points define the enhanced polyhedron $\nabla_{5}$. That means the real F-theory compact four-manifold is obtained.
\begin{figure}[H]
\centering
\includegraphics[scale=0.4]{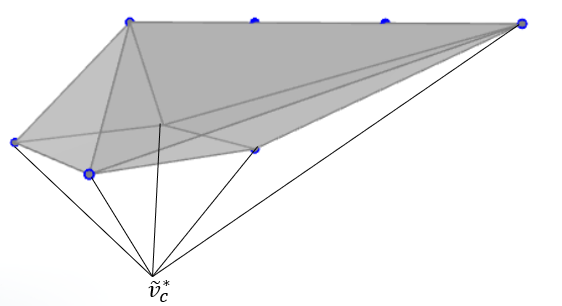}
\caption{Diagram: projected polyhedra of coincident D-brane}
\end{figure}
 The compactifying point is chosen according to: 1, the origin is to be included in the enhanced polyhedron 2, The polyhedron and dual polyhedron are convex polyhedron.
\subsection{The generalized GKZ systems and local solutions}
The periodic integral satisfies the generalized GKZ system, from which the mirror map and the superpotential can be obtained. Combined with toric geometry, we can get the following formula for the differential operator of generalized GKZ system:
\be
\begin{split}
\mathcal{L}(l^a)=&\prod_{k=1}^{l^a_0}(\vartheta_0-k)\prod_{l^a_j>0}\prod_{k=0}^{l^a_j-1}(\vartheta_j-k)-\\
&(-1)^{l^a_0}z_a\prod_{k=1}^{-l^a_0}(\vartheta_0-k)\prod_{l^a_j<0}\prod_{k=0}^{-l^a_j-1}(\vartheta_j-k),
\end{split}
\ee
Where $\vartheta_j=a_j\frac{\partial}{\partial a_j}$ are logarithmic derivative operators of the parameter $a_j$, and $a_j$ defines the coefficients of the polynomial for the hypersurface. $l^a$ are Mori cone generators of toric variety $P_{\Sigma(\nabla_5)}$. These generators are also known as the charge vectors of the gauged linear sigma model (GLSM), $a=1,\ldots,k=h^{1,1}(W_4)$.\\
According to the generators of Mori cone\cite{FS.03,F.01,Wi.93}, the local coordinates near the limit point of large complex structure in $M_4$ moduli space have the following forms:
\be\label{za}
z_a=(-1)^{l_0^a}\prod_j a_j^{l_j^a}.
\ee
Mori Cone is dual to K\"{a}hler cone, and we select a set of base $J_a$ of $H^{1,1}(W_4)$ which is dual to $l^a$. The local coordinates of corresponding K\"{a}hler module space are recorded as $t_a$. The $t_a$ is the local coordinates near the large radius limit point in the $W_4$ K\"{a}hler moduli space, also known as the flat coordinate.\\
Local solution of GKZ system can be derived from basic cycle $w_0$\cite{HLY.96}
\be
w_0(z;\rho)=\sum_{m_1,...,m_k\geq 0}\frac{\Gamma(-\sum (m_k+\rho_k)l_0^a+1)}{\prod_{1\leq i\leq p}\Gamma (\sum(m_k+\rho_k)l_i^a+1)}z^{m+\rho},
\ee
Using the Frobenius method, the complete cycle vector has the following form:
\be
\vec{\Pi}(z)=\begin{pmatrix} \Pi_0=&w_0(z;\rho)|_{\rho=0} \\
\Pi_{1,i}=&\partial_{\rho_i}w_0(z;\rho)|_{\rho=0}\\
\Pi_{2,n}=&\sum_{i_1,i_2}K_{i_1i_2;n}\partial_{\rho_{i_1}}\partial_{\rho_{i_2}}w_0(z;\rho)|_{\rho=0}\\
&\vdots\end{pmatrix},
\ee
$n\subset {1,\cdots,h}$, where $h$ equals the dimension of $H^4(\tilde{M}_4)$. The $K_{i_1i_2;n}$ are the combinatoric coefficients of the second derivative of $w_0$.\\
From this we can obtain the superpotential on B-model and the relative periods which are solutions of Picard-Fuchs equations. Then the corresponding instanton expansion on A-model can be calculated by mirror map :
\be
\vec{\Pi}^*(t)=\begin{pmatrix} \Pi^*_0=&1\\
\Pi^*_{1,i}=&t_i\\
\Pi^*_{2,n}=&\sum_{i_1,i_2}K^*_{i_1i_2;n}t_{i_1}t_{i_2}+b_n+F^{inst}_n\\
&\vdots \end{pmatrix},
\ee
Where $t_i=\Pi_{1,i}/\Pi_0$ are the flat coordinates. The coefficients $K^*_{i_1i_2;n}$ of leading terms is related to the classical sector in superpotential, $K^*_{i_1i_2;n} = K_{i_1i_2;n}$, and $b_n$ are constants. The $F^{inst}_n$ stand for the instanton correction sector of the solutions.\\
mirror map:
\be
t_i=\frac{\Pi_{1,i}(z)}{\Pi_{0}(z)},
\ee
superpotnetial:
\be
W=\sum_{n}C_{n}\Pi_{2,n}(z).
\ee
The combination coefficient $C_n$ in the formula can be determined by matching the leading term of the superpotential in the A-model and B-model. In A-model, the leading term or the classical part can be obtained by calculating the K\"{a}hler volume of the four-dimensional closed-chain $\pi_4\in H_{4}(W_4,Z)$:
\be
\frac{1}{2}\int_{\pi_4}J\wedge J,                 ~~~ J=\sum_{a}t_{a}J{a}.
\ee
The instanton corrections are encoded as a power series expansion of $q_i=exp(2\pi i t_i)$
\be
F^{inst}(t,\hat{t})=\sum_{\vec{r},\vec{m}}G_{\vec{r},\vec{m}}q^{\vec{r}}\hat{q}^{\vec{m}}=\sum_n\sum_{\vec{r},\vec{m}}\frac{N_{\vec{r},\vec{m}}}{n^2}q^{n\vec{r}}\hat{q}^{n\vec{m}}.
\ee
$\{G_{\vec{r},\vec{m}}\}$ are open Gromov-Witten invariants and $\{N_{\vec{r},\vec{m}}\}$ are Ooguri-Vafa invariants.
$\vec{m}$ represent the elements of $H_1(L)$ and $\vec{r}$ represent the elements of $H_2(W_3)$.

\section{Model}
\subsection{D-brane system on the mirror quintic}
The quintic is defined as a hypersurface with polynomial P:
\be
P=a_1x_1^5+a_2x_2^5+a_3x_3^5+a_4x_4^5+a_5x_5+a_0x_1x_2x_3x_4x_5.
\ee
The degree 5 hypersurface P is in the ambient toric variety $P_{\Sigma(\Delta_4)}$. And the toric variety is determined by the vertices of the polyhedron $\Delta_4$:
\be
\begin{split}
&v_1=(4,-1,-1,-1), v_2=(-1,4,-1,-1), \\
&v_3=(-1,-1,4,-1), v_4=(-1,-1,-1,4),\\
&v_5=(-1,-1,-1,-1).
\end{split}
\ee
The vertice of its dual polyhedron $\nabla_4$ are:
\be
\begin{split}
&v_0^*=(0,0,0,0), v_1^*=(1,0,0,0), \\
&v_2^*=(0,1,0,0), v_3^*=(0,0,1,0),\\
&v_4^*=(0,0,0,1), v_5^*=(-1,-1,-1,-1).
\end{split}
\ee
\subsubsection{Parallel phase of three D-branes}
The polyhedron $\nabla_5$ corresponding to the compact 4-fold $W_4$ of F theory compactification consists of vertices $\tilde{\nabla}_5$ (\ref{v3}) and a extra vertex $\tilde{v}^*_c$. However, the detail of the $P^1$ compactification only dominates the subleading term in $g_s$ and would be irrelevant in the decoupling limit. \\
"$\tilde{v}^*_c=(-1,0,0,0,-1)$" is select as the compactifying point.\\
We consider the reducible divisor $\mathcal{D} = \mathcal{D}_1 + \mathcal{D}_2+ \mathcal{D}_3$ defines the parallel D-branes, it can be written as a degree $15$ homogeneous equations:
\begin{alignat}{2}
Q&=b_0(x_1x_2x_3x_4x_5)^3+b_1x_1^7x_2^2x_3^2x_4^2x_5^2+b_2x_1^11+b_3x_1^{15}\nonumber\\
 &\sim\prod_{i=1}^3(\phi_ia_0x_1x_2x_3x_4x_5+a_1x_1^5).
\end{alignat}
The vertices of the enhanced polyhedron $\tilde{\nabla}_5$ for this open and closed system are as follows:
\be
\begin{split}{\label{v3}}
&\tilde{v}_0^*=(0,0,0,0,0), \tilde{v}_1^*=(1,0,0,0,0), \\
&\tilde{v}_2^*=(0,1,0,0,0), \tilde{v}_3^*=(0,0,1,0,0), \\
&\tilde{v}_4^*=(0,0,0,1,0), \tilde{v}_5^*=(-1,-1,-1,-1,0),\\
&\tilde{v}_{6}^*=(0,0,0,0,1),\tilde{v}^*_7=(1,0,0,0,1),\\
&\tilde{v}^*_8=(2,0,0,0,1),\tilde{v}^*_9=(3,0,0,0,1).
\end{split}
\ee
The generators of Mori cone determined by $\nabla_5$ are given:
\begin{equation}{\label{L3}}
\setlength{\arraycolsep}{2pt}
\begin{array}{ccccccccccccc}
       &0 &1&2&3&4&5&6&7&8&9&c& \\
  l^1=(&-2&-2&1&1&1&1&-1&0&0&1&0&)\\
  l^2=(&0&0&0&0&0&0&1&-2&1&0&0&)\\
  l^3=(&0&0&0&0&0&0&0&1&-2&1&0&)\\
  l^4=(&-1&1&0&0&0&0&0&0&1&-1&0&)\\
  l^5=(&0&-2&0&0&0&0&0&0&0&1&1&).
\end{array}
\end{equation}
A suitable set of bases is selected to visualize the closed and open moduli.
\be
t=k_1+k_2+2k_3+3k_4,~~\hat{t}_1=k_2+k_3+k_4,~~\hat{t}_2=k_3+k_4,~~\hat{t}_3=k_4,
\ee
the leading terms of the periods are:
\be\label{leading_term3}
\tilde{\Pi}^*_{2,1}=\frac{5}{2}t^2,~~\tilde{\Pi}^*_{2,2}=2(t-\hat{t}_1)^2,~~\tilde{\Pi}^*_{2,3}=2(t-\hat{t}_2)^2,~~\tilde{\Pi}^*_{2,4}=2(t-\hat{t}_3)^2.
\ee
The $\tilde{\Pi}^*_{2,1}$ that depends on the close moduli $t$ is supposed to be the leading term of the bulk potential function $F_{t}(t)$, while the $\tilde{\Pi}^*_{2,2}$ , $\tilde{\Pi}^*_{2,3}$ and $\tilde{\Pi}^*_{2,4}$  that depend on both open $\hat{t}$ and closed $t$ parameters are supposed to lead the D-brane superpotential $\mathcal{W}_1(t,\hat{t}_1)$, $\mathcal{W}_2(t,\hat{t}_2)$ and $\mathcal{W}_3(t,\hat{t}_3)$.\\
Using algebraic coordinates (\ref{za})
\be
\begin{split}
&z_1=\frac{a_2a_3a_4a_5b_3}{a_0^2a_1^2b_0},~~~z_2=\frac{b_0b_2}{b_1^2},~~~z_3=\frac{b_1b_3}{b_2^2},\\
&z_4=-\frac{a_1b_2}{a_0b_3},
\end{split}
\ee
the fundamental period and the logarithmic periods:
\begin{equation}
\begin{split}
&\Pi_0(z)=w_0(z;0),\\
&\Pi_{1,i}(z)=\partial_{\rho_i}w_0(z;\rho)|_{\rho_i=0},\\
&\Pi_{2,n}(z)=\sum_{i,j}K_{i,j;n}\partial_{\rho_i}\partial_{\rho_j}w_0(z;\rho)|_{\rho=0},
\end{split}
\end{equation}
The flat coordinates are given by
\be
k_i=\frac{\Pi_{1,i}(z)}{\Pi_0(z)}=\frac{1}{2\pi i}log~z_i+...~.
\ee
The mixed inverse mirror maps with $q_i=exp(2\pi ik_i)$ and $\{i=1,2,3,4\}$ are:
\be
\begin{split}
z_1=&{q_1}-{q_1}{q_4}-{q_1}{q_3}{q_4}-{q_1}{q_2}{q_3}{q_4}+{q_1}{q_3}{q_4}^2 \\
&+24{q_1}^2{q_3}{q_4}^2 +{q_1}{q_2}{q_3}{q_4}^2+24{q_1}^2{q_2}{q_3}{q_4}^2+ ...\\
z_2=&{q_2}-2q_2^2+{q_2}{q_3} + 3q_2^3-3{q_2}^2{q_3}-2{q_2}^2{q_3}^2 \\
&-2{q_2}{q_3}{q_4} + 8{q_2}^2{q_3}{q_4}+...\\
z_3=&{q_3}+{q_2}{q_3} -2q_3^2 -3{q_2}{q_3}^2 +3{q_3}^3 +7{q_2}{q_3}^3 \\
&-2{q_2}^2{q_3}^2+...\\
z_4=&{q_4}+{q_3}{q_4} +q_4^2 +{q_2}{q_3}{q_4} +{q_3}^2{q_4}^2 +{q_4}^3\\
& +{q_2}^2{q_3}^2{q_4}^2+...~.
\end{split}
\ee
According to the leading terms (\ref{leading_term3}), the relative periods which corresponds to the closed-string period and D-brane superpotentials in the A-model as follows:

\be
\begin{split}
F_{t}(t)\equiv\Pi_{2,1}=&\frac{5}{2}t^2+\frac{1}{4\pi^2}(2875q+\frac{4876875}{4}q^2+...),\\
\mathcal{W}_1(t,\hat{t}_1)\equiv\Pi_{2,2}=&2(t-\hat{t}_1)^2+\frac{2}{4\pi^2}( 800q + 340000q^2\\
&-160q\hat{q}_1^{-1} +6600q^2\hat{q}_{1}^{-2}+10\hat{q}_{1}\\
&-58280q^2\hat{q}_{1}^{-1}+\frac{{5\hat{q}_{1}^{2}}}{2}\\
&+1020q\hat{q}_{1} +\frac{{10\hat{q}_1^3}}{9} +...),\nonumber
\end{split}
\ee
\be
\begin{split}
\mathcal{W}_2(t,\hat{t}_2)\equiv\Pi_{2,3}=&2(t-\hat{t}_2)^2+\frac{2}{4\pi^2}( 800q + 340000q^2 \\
&-160q\hat{q}_{2}^{-1} +6600q^2\hat{q}_{2}^{-2} + 10\hat{q}_{2} \\
&-58280q^2\hat{q}_{2}^{-1}+\frac{{5\hat{q}_{2}^{2}}}{2}\\
&+1020q\hat{q}_{2} +\frac{{10\hat{q}_2^3}}{9} +...),\\
\mathcal{W}_3(t,\hat{t}_3)\equiv\Pi_{2,4}=&2(t-\hat{t}_3)^2+\frac{2}{4\pi^2}( 800q + 340000q^2 \\
&-160q\hat{q}_{3}^{-1} + 6600q^2\hat{q}_{3}^{-2} + 10\hat{q}_{3} \\
&-58280q^2\hat{q}_{3}^{-1}+\frac{{5\hat{q}_{3}^{2}}}{2}\\
&+ 1020q\hat{q}_{3} +\frac{{10\hat{q}_3^3}}{9}  +...).
\end{split}
\ee
The disk invariants are Table \ref{tab:3}.
\begin{center}
\footnotesize
\def\temptablewidth{0.35\textwidth}
\begin{tabular*}{\temptablewidth}{c@{\extracolsep{\fill}}|ccccc}
 \hline $n_1=n_3 \backslash n_2=0,n_4~~  $&1&2&3&4  \\ \hline
 1& 0&-320&0& 0   \\
 2& 0&0&0& 13280  \\
\end{tabular*}
\tabcaption{\label{tab:3} \footnotesize  Ooguri-Vafa invariants
$N_{n_1,n_2,n_3,n_4}$ from the off-shell superpotential $\mathcal{W}_1(t,\hat{t})$ contributed by one of three parallel branes on the mirror quintic.}
\end{center}
The D-brane superpotential with one open deformation modulus defined by the divisor $Q=(\phi_1a_0x_1x_2x_3x_4x_5+a_1x_1^5)$ and the two parallel D-branes superpotentials defined by the divisor $Q=b_0(x_1x_2x_3x_4x_5)^2+b_1x_1^6x_2x_3x_4x_5+b_2x_1^{10}\sim\prod_{i=1}^2(\phi_ia_0x_1x_2x_3x_4x_5+a_1x_1^5)$ are the same as reference\cite{AHJMMS.09}. And so do the Ooguri-Vafa invariants.
\subsubsection{Complete coincident phase of three D-branes}
The coincidence can be two or all of them. As we know, the parallel D-branes geometry corresponds to the Coulomb branch of the gauge theory on the worldvolume. When parallel D-brane coincide, the gauge group $U(1) \times U(1) \times U(1)$ is enhanced to the gauge group $U(3)$, and the Coulomb branch transform to the Higgs branch of the gauge theory. It's easy to deduce that three parallel branes is coincident, but two of the three parallel branes is coincident, which is more complicated, it present part coincident phase $U(2) \times U(1)$. Firstly, the result of complete coincidence (Higgs branch) of three parallel branes is listed here. \\
We ignore the interior point $\tilde{v}^*_7$ and $\tilde{v}^*_8$, and select $\tilde{v}^*_c=(-1,0,0,0,-1)$ as the compactifying point.\\
The generators of Mori cone determined by $\nabla_5$ are given:
\begin{equation}
\setlength{\arraycolsep}{2pt}
\begin{array}{ccccccccccccccccccccccccc}
          & 0 &1&2&3&4&5&6&7&c & \\
  l^1=( &-2&-2&1&1&1&1&-1&1&0 &)\\
  l^2=( &-3&3&0&0&0&0&1&-1&0 &)\\
  l^3=( &0&-2&0&0&0&0&0&1&1 &).
\end{array}
\end{equation}
A suitable set of bases is selected to visualize the closed and open moduli.
\be
t=k_1+k_2,~~\hat{t}=k_2,
\ee
the relative periods which corresponds to the closed-string period and D-brane superpotentials in the A-model as follows:
\be\
\begin{split}
F_{t}(t)\equiv\Pi_{2,1}=&\frac{5}{2}t^2+\frac{1}{4\pi^2}(2875q+\frac{4876875}{4}q^2+\\
&\frac{8564575000}{9}q^3+...),\\
\mathcal{W}(t,\hat{t})\equiv\Pi_{2,2}=&2(t-\hat{t})^2+\frac{1}{4\pi^2}(2580q + 866769q^2 +\\
&\frac{1879614800}{3}q^3 -24q\hat{q}^{-1}+ 54q^2\hat{q}^{-2} + 4\hat{q} \\
&+13944q^2\hat{q}^{-1}+\hat{q}^2- 6552q\hat{q}+\\
&10012416q^3\hat{q}^{-1} +\frac{4\hat{q}^3}{9}+5940q\hat{q}^2 +...).
\end{split}
\ee
The disk invariants are Table \ref{tab:9}.
\begin{table*}
\begin{center}
\def\temptablewidth{1.0\textwidth}
\begin{tabular*}{\temptablewidth}{@{\extracolsep{\fill}}c|ccccc}
 \hline $n_2\backslash n_1$&0&1&2&3&4  \\ \hline
 0& 0& -24& 60& -264& 1824 \\
 1& 4& 2580& 139440& -42304& 282600  \\
 2& 0& -6552& 866124& 10012416& 6266256  \\
 3& 0& 5940& -11124972& 626537980& 7877324052  \\
 4& 0& -1944& 46600344& -14170127808& 618549866964 \\
\end{tabular*}
\tabcolsep 0pt \caption{\label{tab:9}  Ooguri-Vafa invariants
$N_{n_1,n_2}$ from the off-shell superpotential $\mathcal{W}(t,\hat{t})$ contributed by the complete coincident phase of three D-branes on the mirror quintic.} \vspace*{-12pt}
\end{center}
\end{table*}

\subsubsection{Part coincident D-branes phase}
"$\tilde{v}^*_c=(-1,0,0,0,-1)$" is select as the compactifying point. \\
First, we ignore the interior point $\tilde{v}^*_7$ on the one-dimensional edge with $\tilde{v}_6^*$ , $\tilde{v}_8^*$ and $\tilde{v}_9^*$ to obtain the new charge vectors.
The generators of Mori cone determined by $\nabla_5$ are given:
\begin{equation}
\setlength{\arraycolsep}{2pt}
\begin{array}{ccccccccccccccccccccccccc}
        & 0 &1&2&3&4&5&6&7&8&c & \\
  l^1=( &-2&-2&1&1&1&1&-1&0&1&0 &)\\
  l^2=( &0&0&0&0&0&0&1&-3&2&0 &)\\
  l^3=( &-1&1&0&0&0&0&0&1&-1&0 &)\\
  l^4=( &0&-2&0&0&0&0&0&0&1&1 &).
\end{array}
\end{equation}
A suitable set of bases is selected to visualize the closed and open moduli.
\be
t_1=k_1+k_2+3k_3,~~\hat{t}_1=k_2+k_3,~~\hat{t}_2=k_3.
\ee
D-brane superpotentials in the A-model as follows:
\be
\begin{split}
F_{t}(t)\equiv\Pi_{2,1}=&\frac{5}{2}t^2+\frac{1}{4\pi^2}(2875q+...),\\
\mathcal{W}_1(t,\hat{t}_1)\equiv\Pi_{2,2}=&2(t-\hat{t}_1)^2+\frac{2}{4\pi^2}( 704q +24q\hat{q}_1^{-1}-4\hat{q}_{1} \\
&-5\hat{q}_{1}^{2}+648q\hat{q}_{1}+\frac{92\hat{q}_{1}^{3}}{9}...),\\
\mathcal{W}_2(t,\hat{t}_2)\equiv\Pi_{2,3}=&2(t-\hat{t}_2)^2+\frac{2}{4\pi^2}( 800q -160q\hat{q}_{2}^{-1} +\\
&6600q^2\hat{q}_{2}^{-2} +10\hat{q}_{2}+\frac{5\hat{q}_{2}^{2}}{2}+\\
&1020q\hat{q}_{2} +\frac{10\hat{q}_2^3}{9}+...).\\
\end{split}
\ee
Second, we ignore the interior point $\tilde{v}^*_8$ on the one-dimensional edge with $\tilde{v}_6^*$ , $\tilde{v}_7^*$ and $\tilde{v}_9^*$ to obtain the new charge vectors.
The generators of Mori cone determined by $\nabla_5$ are given:
\begin{equation}
\setlength{\arraycolsep}{2pt}
\begin{array}{ccccccccccccccccccccccccc}
        & 0 &1&2&3&4&5&6&7&8&c & \\
  l^1=(&-2&-2&1&1&1&1&-1&0&1&0 &)\\
  l^2=(&0&0&0&0&0&0&2&-3&1&0 &)\\
  l^3=(&-2&2&0&0&0&0&0&1&-1&0 &)\\
  l^4=(&0&-2&0&0&0&0&0&0&1&1 &).
\end{array}
\end{equation}
A suitable set of bases is selected to visualize the closed and open moduli.
\be
t_1=2k_1+k_2+3k_3,~~\hat{t}_1=k_2+k_3,~~\hat{t}_2=k_3.
\ee
D-brane superpotentials in the A-model as follows:
\be
\begin{split}
F_{t}(t)\equiv\Pi_{2,1}=&\frac{5}{2}t^2+\frac{1}{4\pi^2}(21040875q+...),\\
\mathcal{W}_1(t,\hat{t}_1)\equiv\Pi_{2,2}=&2(t-\hat{t}_1)^2+\frac{2}{4\pi^2}( 5653152q +\\
&26400q\hat{q}_1^{-1}-26\hat{q}_{1} +\frac{47\hat{q}_{1}^{2}}{6}-\\
&14885404q\hat{q}_{1}-\frac{19601920\hat{q}_{1}^{3}}{9}...),\\
\mathcal{W}_2(t,\hat{t}_2)\equiv\Pi_{2,3}=&2(t-\hat{t}_2)^2+\frac{2}{4\pi^2}( 5995056q -\\
&56568q\hat{q}_{2}^{-1} +26\hat{q}_{2}+\frac{47\hat{q}_{2}^{2}}{6}+\\
&19704132q\hat{q}_{2} +\frac{178\hat{q}_2^3}{45}+...).\\
\end{split}
\ee
\subsection{D-brane system on hypersurface \\ $X_8(1,1,2,2,2)$}
The hypersurface is defined as the polynomial P:
\be
\begin{split}
P=&a_1x_1^8+a_2x_2^8+a_3x_3^4+a_4x_4^4+a_5x_5^4+\\
&a_0x_1x_2x_3x_4x_5+a_6x_1^4x_2^4.
\end{split}
\ee
The degree 8 hypersurface P is in the ambient toric variety $P_{\Sigma(\Delta_4)}$. And the toric variety is determined by the vertices of the polyhedron $\Delta_4$:
\be
\begin{split}
&v_1=(-1,-1,-1,-1), v_2=(7,-1,-1,-1),\\
&v_3=(-1,3,-1,-1),v_4=(-1,-1,3,-1),\\
&v_5=(-1,-1,-1,3).
\end{split}
\ee
The vertice of its dual polyhedra $\nabla_4$ are:
\be
\begin{split}
&v_0^*=(0,0,0,0),v_1^*=(-1,-2,-2,-2),\\
&v_2^*=(1,0,0,0), v_3^*=(0,1,0,0),\\
&v_4^*=(0,0,1,0),v_5^*=(0,0,0,1),\\
&v_6^*=(0,-1,-1,-1).
\end{split}
\ee
\subsubsection{Parallel phase of three D-branes}
"$\tilde{v}^*_c=(0,-1,0,0,-1)$" is select as the compactifying point. \\
We consider the reducible divisor $\mathcal{D} = \mathcal{D}_1 + \mathcal{D}_2+ \mathcal{D}_3$ defines the parallel D-branes, it can be written as a degree $15$ homogeneous equations:
\begin{alignat}{2}
\begin{split}
Q&=b_0(x_1x_2x_3x_4x_5)^3+b_1x_3^6x_1^2x_2^2x_4^2x_5^2+b_2x_3^9x_1x_2x_4x_5+b_3x_1^{12}\\
 &\sim\prod_{i=1}^3(\phi_ia_0x_1x_2x_3x_4x_5+a_1x_1^4).
\end{split}
\end{alignat}
The vertices of the enhanced polyhedron $\tilde{\nabla}_5$ for this open and closed system are as follows:
\be
\begin{split}{\label{V3}}
&\tilde{v}_0^*=(0,0,0,0,0), \tilde{v}_1^*=(-1,-2,-2,-2,0), \\
&\tilde{v}_2^*=(1,0,0,0,0),\tilde{v}_3^*=(0,1,0,0,0),\\
&\tilde{v}_4^*=(0,0,1,0,0), \tilde{v}_5^*=(0,0,0,1,0),\\
&\tilde{v}_{6}^*=(0,-1,-1,-1,0),\tilde{v}^*_7=(0,0,0,0,1),\\
&\tilde{v}^*_8=(0,1,0,0,1),\tilde{v}^*_9=(0,2,0,0,1),\\
&\tilde{v}^*_{10}=(0,3,0,0,1).
\end{split}
\ee
The generators of Mori cone determined by $\nabla_5$ are given:
\begin{equation} \label{LL}
\setlength{\arraycolsep}{2.0pt}
\begin{array}{ccccccccccccccccccccccccc}
        & 0 &1&2&3&4&5&6&7&8&9&10&c & \\
  l^1=( &-1&0&0&-2&1&1&1&-1&0&0&1&0 &)\\
  l^2=( &0&1&1&0&0&0&-2&0&0&0&0&0 &)\\
  l^3=( &0&0&0&0&0&0&0&1&-2&1&0&0 &)\\
  l^4=( &0&0&0&0&0&0&0&0&1&-2&1&0 &)\\
  l^5=( &-1&0&0&1&0&0&0&0&0&1&-1&0 &)\\
  l^6=( &0&0&0&-2&0&0&0&0&0&0&1&1 &).
\end{array}
\end{equation}
A suitable set of bases is selected to visualize the closed and open moduli.
\be
\begin{split}
&t_1=k_1+k_3+2k_4+3k_5,~~t_2=k_2,\\
&\hat{t}_1=k_3+k_4+k_5,~~\hat{t}_2=k_4+k_5,~~\hat{t}_3=k_5.
\end{split}
\ee
the leading terms of the periods are:
\be\label{leading_termLL}
\begin{split}
&\tilde{\Pi}^*_{2,1}=2t_1^2,~~\tilde{\Pi}^*_{2,2}=\frac{3}{2}(t_1-\hat{t}_1)^2,~~\tilde{\Pi}^*_{2,3}=\frac{3}{2}(t_1-\hat{t}_2)^2,\\
&\tilde{\Pi}^*_{2,4}=\frac{3}{2}(t_1-\hat{t}_3)^2.
\end{split}
\ee
The $\tilde{\Pi}^*_{2,1}$ that depends on the close moduli $t$ is supposed to be the leading term of the bulk potential function $F_{t}(t)$, while the $\tilde{\Pi}^*_{2,2}$ , $\tilde{\Pi}^*_{2,3}$ and $\tilde{\Pi}^*_{2,4}$  that depend on both open $\hat{t}$ and closed $t$ parameters are supposed to lead the D-brane superpotential $\mathcal{W}_1(t,\hat{t}_1)$, $\mathcal{W}_2(t,\hat{t}_2)$ and $\mathcal{W}_3(t,\hat{t}_3)$.\\
Using algebraic coordinates (\ref{za})
\be
\begin{split}
&z_1=-\frac{a_4a_5a_6b_3}{a_0a_3^2b_0},~~~z_2=\frac{a_1a_2}{a_6^2},~~~z_3=\frac{b_0b_2}{b_1^2},\\
&z_4=\frac{b_1b_3}{b_2^2},~~~z_5=-\frac{a_3b_2}{a_0b_3},
\end{split}
\ee
the fundamental period and the logarithmic periods:
\begin{equation}
\begin{split}
&\Pi_0(z)=w_0(z;0),\\
&\Pi_{1,i}(z)=\partial_{\rho_i}w_0(z;\rho)|_{\rho_i=0},\\
&\Pi_{2,n}(z)=\sum_{i,j}K_{i,j;n}\partial_{\rho_i}\partial_{\rho_j}w_0(z;\rho)|_{\rho=0},
\end{split}
\end{equation}
The flat coordinates are given by
\be
k_i=\frac{\Pi_{1,i}(z)}{\Pi_0(z)}=\frac{1}{2\pi i}log~z_i+...~.
\ee
The mixed inverse mirror maps with $q_i=exp(2\pi ik_i)$ and $\{i=1,2,3,4,5\}$ are:
\be
\begin{split}
z_1=&{q_1}+{q_1}{q_2}-{q_1}{q_5}-{q_1}{q_2}{q_5}-{q_1}{q_4}{q_5}-{q_1}{q_2}{q_4}{q_5}-\\
&{q_1}{q_3}{q_4}{q_5}-{q_1}{q_2}{q_3}{q_4}{q_5}+{q_1}{q_4}{q_5}^2+6{q_1}^2{q_4}{q_5}^2+\\ &{q_1}{q_2}{q_4}{q_5}^2+12{q_1}^2{q_2}{q_4}{q_5}^2+6{q_1}^2{q_2}^2{q_4}{q_5}^2+ ...\\
z_2=&{q_2}-2q_2^2+3{q_2}^3-48{q_1}{q_2}{q_3}{q_4}^2{q_5}^3+240{q_1}{q_2}^2{q_3}{q_4}^2{q_5}^3\\
&-624{q_1}{q_2}^3{q_3}{q_4}^2{q_5}^3+...\\
z_3=&{q_3}+{q_3}{q_4} -2{q_3}^2 -3{q_3}^2{q_4} -2{q_3}^2{q_4}^2 -2{q_3}{q_4}{q_5} +\\
&8{q_3}^2{q_4}{q_5}+8{q_3}^2{q_4}^2{q_5}+...\\
z_4=&{q_4}+{q_3}{q_4} -2q_4^2 -3{q_3}{q_4}^2 -2{q_3}^2{q_4}^2 -{q_4}{q_5} -{q_3}{q_4}{q_5}\\
&+5{q_4}^2{q_5}+6{q_3}{q_4}^2{q_5}+5{q_3}^2{q_4}^2{q_5}+...\\
z_5=&{q_5}+{q_4}{q_5} +q_5^2 +{q_3}{q_4}{q_5} +{q_4}^2{q_5}^2  +{q_3}^2{q_4}^2{q_5}^2+\\
&{q_5}^3-18{q_1}{q_3}{q_4}^2{q_5}^3+...~.
\end{split}
\ee
According to the leading terms (\ref{leading_termLL}), the relative periods which corresponds to the closed-string period and D-brane superpotentials in the A-model as follows:

\begin{equation*}
\begin{split}
F_{t}(t)\equiv\Pi_{2,1}=&2t_1^2+\frac{1}{4\pi^2}(4q_2+q_2^2+\frac{4}{9}q_2^3+640q_1q_2+\\
&q_1^2(72224q_2+20224q_2^2)+...),\\
\mathcal{W}_1(t,\hat{t}_1)\equiv\Pi_{2,2}=&\frac{3}{2}(t_1-\hat{t}_1)^2+\frac{1}{4\pi^2}( 3q_2 + \frac{3}{4}q_2^2 +12\hat{q}_1+\\
&3\hat{q}_2^2-27q_1\hat{q}_1^{-1}+ 66q_1\hat{q}_1-63q_1q_2\hat{q}_1^{-1} +\\
&360q_1q_2 + 282\hat{q}_{1}q_1q_2 -\frac{405}{4}q_1^2\hat{q}_{1}^{-2}-\\
&324q_1^2\hat{q}_{1}^{-1}+...),\\
\end{split}
\end{equation*}
\begin{equation*}
\begin{split}
\mathcal{W}_2(t,\hat{t}_2)\equiv\Pi_{2,3}=&\frac{3}{2}(t_1-\hat{t}_2)^2+\frac{1}{4\pi^2}( 3q_2 + \frac{3}{4}q_2^2 +12\hat{q}_2+\\
&3\hat{q}_2^2-27q_1\hat{q}_2^{-1}+ 66q_1\hat{q}_2-63q_1q_2\hat{q}_2^{-1} +\\
&360q_1q_2 + 282\hat{q}_{2}q_1q_2 -\frac{405}{4}q_1^2\hat{q}_{2}^{-2}-\\
&324q_1^2\hat{q}_{2}^{-1}+...),
\end{split}
\end{equation*}
\begin{equation}\label{sppLL}
\begin{split}
\mathcal{W}_3(t,\hat{t}_3)\equiv\Pi_{2,3}=&\frac{3}{2}(t_1-\hat{t}_3)^2+\frac{1}{4\pi^2}( 3q_2 + \frac{3}{4}q_2^2 +12\hat{q}_3+\\
&3\hat{q}_3^2-27q_1\hat{q}_3^{-1}+ 66q_1\hat{q}_3-63q_1q_2\hat{q}_3^{-1} +\\
&360q_1q_2 + 282\hat{q}_{3}q_1q_2 -\frac{405}{4}q_1^2\hat{q}_{3}^{-2}-\\
&324q_1^2\hat{q}_{3}^{-1}+...).
\end{split}
\end{equation}
The disk invariants are Table \ref{tab:13}.
\begin{center}
\footnotesize
\label{table210}
\def\temptablewidth{0.4\textwidth}
\begin{tabular*}{\temptablewidth}{c@{\extracolsep{\fill}}|cccccc}
 \hline $n_1=n_2=n_3=0~~n_5\backslash n_4$&0&1&2  \\ \hline
 0& 0& 0& 0&  \\
 1& 0& 12& 0&  \\
 2& 0& 0& 0&  \\
 \end{tabular*}
 \end{center}
\begin{center}
\footnotesize
\def\temptablewidth{0.4\textwidth}
\begin{tabular*}{\temptablewidth}{c@{\extracolsep{\fill}}|cccccc}
 \hline $n_1=0,n_2=n_3=1~~n_5\backslash n_4$&0&1&2  \\ \hline
 0& 3& 0& 0&  \\
 1& 0& 0& 0&  \\
 2& 0& 0& 0& \\
\end{tabular*}
\tabcaption{ \label{tab:13} \footnotesize  Ooguri-Vafa invariants $N_{n_1,n_2,n_3,n_4,n_5}$ from the off-shell superpotential $\mathcal{W}_1(t,\hat{t})$ contributed by one of three parallel branes on the hypersurface $X_8(1,1,2,2,2)$.}
\end{center}
The D-brane superpotential with one open deformation modulus defined by the divisor $Q=b_0(x_1x_2x_3x_4x_5)+b_1x_3^4$
and two parallel D-branes superpotentials defined by the divisor $Q=b_0(x_1x_2x_3x_4x_5)^2+b_1x_3^5x_1x_2x_4x_5+b_2x_3^8\sim\prod_{i=1}^2(\phi_ia_0x_1x_2x_3x_4x_5+a_3x_3^4)$ are the same as reference\cite{JXT.17}.
\subsubsection{Complete coincident phase of three D-branes}
We ignore the interior point $\tilde{v}^*_8$ and $\tilde{v}^*_9$, and select $\tilde{v}^*_c=(0,-1,0,0,-1)$ as the compactifying point.\\
The generators of Mori cone determined by $\nabla_5$ are given:
\begin{equation}
\setlength{\arraycolsep}{2.0pt}
\begin{array}{ccccccccccccccccccccccccc}
     & &  & 0 &1&2&3&4&5&6&7&8&c & \\
  l^1&=&( &-1&0&0&-2&1&1&1&-1&1&0 &)\\
  l^2&=&( &0&1&1&0&0&0&-2&0&0&0 &)\\
  l^3&=&( &-3&0&0&3&0&0&0&1&-1&0 &)\\
  l^4&=&( &0&0&0&-2&0&0&0&0&1&1 &).
\end{array}
\end{equation}
A suitable set of bases is selected to visualize the closed and open moduli.
\be
t_1=k_1+k_3,~~t_2=k_2,~~\hat{t}=k_3.
\ee
D-brane superpotentials in the A-model as follows:
\be
\begin{split}
F_{t}(t)\equiv\Pi_{2,1}=&2t_1^2+\frac{1}{4\pi^2}(4q_2+q_2^2+\frac{4}{9}q_2^3+640q_1q_2+\\
&q_1^2(72224q_2+20224q_2^2)+q_1^3(753920q_2+\\
&15078400q_2^2+\frac{7787008}{9}q_2^3)+...),\\
\mathcal{W}(t,\hat{t})\equiv\Pi_{2,2}=&\frac{3}{2}(t_1-\hat{t})^2+\frac{1}{4\pi^2}(3q_2 + \frac{3}{4}q_2^2+3q_1\hat{q}_1^{-1}+\\
&3q_1q_2\hat{q}^{-1}+\frac{3}{4}q_1^2\hat{q}^{-2} +\frac{3q_2^2q_1^2\hat{q}^{-2}}{4} +...).
\end{split}
\ee
The disk invariants are Table \ref{tab:15}.
\begin{center}
\footnotesize
\def\temptablewidth{0.4\textwidth}
\begin{tabular*}{\temptablewidth}{@{\extracolsep{\fill}}lllll}
 \hline   \\
 $N_{0,1,0} = $ 3& $N_{1,0,0} = $ 3& $N_{2,1,0} = $ 3& $N_{3,1,0} = $ 3  \\
 $N_{4,2,0} = $ 6& $N_{1,1,0} = $ 3& $N_{4,3,0} = $ 3& $N_{3,2,0} = $ 3  \\
                 &                 &                 &                   \\
 \hline   \\
 \end{tabular*}
\tabcaption{\label{tab:15} \footnotesize  Ooguri-Vafa invariants
$N_{n_1,n_2,n_3}$ from the off-shell superpotential $\mathcal{W}_1(t,\hat{t})$ contributed by the complete coincident phase of three D-branes on the hypersurface $X_8(1,1,2,2,2)$.}
\end{center}
\subsubsection{Part coincident D-branes phase}
"$\tilde{v}^*_c=(-1,0,0,0,-1)$" is select as the compactifying point.\\
First, we ignore the interior point $\tilde{v}^*_8$ on the one-dimensional edge with $\tilde{v}_7^*$ , $\tilde{v}_9^*$ and $\tilde{v}_{10}^*$ to obtain the new charge vectors.
The generators of Mori cone determined by $\nabla_5$ are given:
\begin{equation}
\setlength{\arraycolsep}{2.0pt}
\begin{array}{ccccccccccccccccccccccccc}
     & &  & 0 &1&2&3&4&5&6&7&8&9&c & \\
  l^1&=&( &-1&0&0&-2&1&1&1&-1&0&1&0 &)\\
  l^2&=&( &0&1&1&0&0&0&-2&0&0&0&0 &)\\
  l^3&=&( &0&0&0&0&0&0&0&1&-3&2&0 &)\\
  l^4&=&( &-1&0&0&1&0&0&0&0&1&-1&0 &)\\
  l^5&=&( &0&0&0&-2&0&0&0&0&0&1&1 &).
\end{array}
\end{equation}
A suitable set of bases is selected to visualize the closed and open moduli.
\be
t_1=k_1+k_3+3k_4,~~t_2=k_2,~~\hat{t}_1=k_3+k_4,~~\hat{t}_2=k_4.
\ee
D-brane superpotentials in the A-model as follows:
\be
\begin{split}
F_{t}(t)\equiv\Pi_{2,1}=&2t_1^2+\frac{1}{4\pi^2}(4q_2+q_2^2+\frac{4}{9}q_2^3+16q_1q_2+...),\\
\mathcal{W}_1(t,\hat{t}_1)\equiv\Pi_{2,2}=&\frac{3}{2}(t_1-\hat{t}_1)^2+\frac{1}{4\pi^2}( 3q_2 + \frac{3}{4}q_2^2 +24q_{1}-\\
&18\hat{q}_{1}-\frac{45\hat{q}_{1}^2}{2}+84q_{1}\hat{q}_1-36q_1q_2\hat{q}_1^{-1}+...),\\
\mathcal{W}_2(t,\hat{t}_2)\equiv\Pi_{2,3}=&\frac{3}{2}(t_1-\hat{t}_2)^2+\frac{1}{4\pi^2}( 3q_2 + \frac{3}{4}q_2^2+18q_1\hat{q}_2^{-1}+\\
&18q_1q_2\hat{q}_2^{-1}  -84q_1^2\hat{q}_{2}^{-3}-224q_{2}q_{1}^{2}\hat{q}_{2}^{-3} +...).
\end{split}
\ee
Second, we ignore the interior point $\tilde{v}^*_9$ on the one-dimensional edge with $\tilde{v}_7^*$ , $\tilde{v}_8^*$ and $\tilde{v}_{10}^*$ to obtain the new charge vectors.
The generators of Mori cone determined by $\nabla_5$ are given:
\begin{equation}
\setlength{\arraycolsep}{2.0pt}
\begin{array}{ccccccccccccccccccccccccc}
     & &  & 0 &1&2&3&4&5&6&7&8&9&c & \\
  l^1&=&( &-1&0&0&-2&1&1&1&-1&0&1&0 &)\\
  l^2&=&( &0&1&1&0&0&0&-2&0&0&0&0 &)\\
  l^3&=&( &0&0&0&0&0&0&0&1&-3&2&0 &)\\
  l^4&=&( &-1&0&0&1&0&0&0&0&1&-1&0 &)\\
  l^5&=&( &0&0&0&-2&0&0&0&0&0&1&1 &).
\end{array}
\end{equation}
A suitable set of bases is selected to visualize the closed and open moduli.
\be
t_1=2k_1+k_3+3k_4,~~t_2=k_2,~~\hat{t}_1=k_3+k_4,~~\hat{t}_2=k_4.
\ee
D-brane superpotentials in the A-model as follows:
\begin{equation*}
\begin{split}
F_{t}(t)\equiv\Pi_{2,1}=&2t_1^2+\frac{1}{4\pi^2}(16q_2+4q_2^2+\frac{16}{9}q_2^3+\\
&112q_1+448q_1q_2+...),\\
\mathcal{W}_1(t,\hat{t}_1)\equiv\Pi_{2,2}=&\frac{3}{2}(t_1-\hat{t}_1)^2+\frac{1}{4\pi^2}( 12q_2 + 3q_2^2 -\\
&\frac{213\hat{q}_1^2}{8}+ 153q_1\hat{q}_1+ 504q_1q_2...),\\
\end{split}
\end{equation*}
\begin{equation*}
\begin{split}
\mathcal{W}_2(t,\hat{t}_2)\equiv\Pi_{2,3}=&\frac{3}{2}(t_1-\hat{t}_2)^2+\frac{1}{4\pi^2}( 12q_2 + 3q_2^2+\\
&42q_1 + 168q_1q_2 +840q_{1}^{2}q_{2}\hat{q}_2^{-4}...).\\
\end{split}
\end{equation*}

\section{Summary}
The effective D-brane superpotential is of great significance to both physics and mathematics. In type II/F-theory compatification, the vacuum structure is determined by the superpotentials, whose second derivative gives the chiral ring structure, and which is the generating function of the Ooguri-Vafa invariants. Those invariants are closely related to the number of the BPS states, which count the holomorphic disks on Calabi-Yau manifolds mathematically. \\
In this paper, for the system with three D-branes on the mirror quintic and hypersurface $X_8(1,1,2,2,2)$, the off-shell effective superpotentials and relevant geometric invariants are calculated by type II/F-theory duality, open-closed mirror symmetry and GKZ-system method. \\
The results show that there are three phases: parallel phase, part coincident phase and complete coincident phase, i.e., the Coulomb branch, the mixed Coulomb-Higgs branch and the Higgs branch, respectively. The phase transitions: the parallel phase $\to$ the part coincident phase $\to$ the complete coincident phase, correspond to the enhancement of gauge group $U(1)\times U(1)\times U(1) \rightarrow U(1)\times U(2) \rightarrow U(3)$ in the low energy effective theory. In the parallel phase, the effective superpotential contributed by one of the three D-branes is the same as that of the D-brane system with single brane in the same Calabi-Yau manifold.  In the part coincident phase, i.e., the two of the three parallel D-branes approach to each other and melt into one, the superpotential and the invariant are different from either those of the system with three D-branes or those of the system with two D-branes. It gives the sign of the phase transition from the $U(1)\times U(1)\times U(1)$ to $U(1)\times U(2)$ in terms of gauge theory. Similarly, in the complete coincident D-brane phase, i.e., the all parallel D-branes approach to each other and melt into one, our calculations indicate that the effective superpotentials and the Ooguri-Vafa invariants are completely different from those of the system with one D-brane, although there is no difference between the coincident D-branes and the single D-brane from the set theory view point. It shows the feature of the phase transition from $U(1)\times U(1)\times U(1)$ to $U(3)$ in terms of gauge theory. Thus, there are various phase structures in the low energy effective theory of the system with three D-branes on the compact Calabi-Yau manifold.\\
Furthermore, we are going to study the more general physical and mathematical properties of D-brane system on compact Calabi-Yau manifolds, especially in phenomenological applications of superpotential and the relevant geometric invariants.\\

\section*{Acknowledgemnts}
This work has been supported by NSFC(11475178) and Y4JT01VJ01.
\end{multicols}

\vspace{10mm}

\begin{multicols}{2}

\end{multicols}

\clearpage

\end{document}